\documentclass[12pt]{article}
\usepackage{graphicx}
\usepackage{lineno} 
\usepackage{subcaption}
\usepackage{mathtools}

\newcommand\pubdate{\today}

\def\clermont{LPC - Clermont Ferrand, CNRS, France\\
}
\def\support{\footnote{On behalf of the LHCb Collaboration.}}

\def\Title#1{\begin{center} {\Large #1 } \end{center}}
\def\Author#1{\begin{center}{ \sc #1} \end{center}}
\def\Address#1{\begin{center}{ \it #1} \end{center}}

\newcommand\pubblock{\rightline{\begin{tabular}{l} \\
         \pubdate  \end{tabular}}}
\newenvironment{Abstract}{\begin{quotation}  }{\end{quotation}}
\newenvironment{Presented}{\begin{quotation} \begin{center} 
             PRESENTED AT\end{center}\bigskip 
      \begin{center}\begin{large}}{\end{large}\end{center} \end{quotation}}

\def\beq{\begin{equation}}
\def\eeq#1{\label{#1}\end{equation}}
\def\eeqn{\end{equation}}

\def\beqa{\begin{eqnarray}}
\def\eeqa#1{\label{#1}\end{eqnarray}}
\def\eeqan{\end{eqnarray}}

\let\bar=\overbar

\def\Dslash{\not{\hbox{\kern-4pt $D$}}}
\def\dslash{\not{\hbox{\kern-2pt $\del$}}}

\def\msb{{\bar{\ssstyle M \kern -1pt S}}}

\newcommand\mBtokpi{B^0 \rightarrow K^+\pi^-}

\newcommand\mantiBtokpi{\bar{B}^0 \rightarrow K^-\pi^+}

\newcommand\mBstokpi{B_s^0 \rightarrow \pi^+ K^-}

\newcommand\Btokpipm{$B^0 \rightarrow K^\pm \pi^\mp$}
\newcommand\Bstokpipm{$B_s^0 \rightarrow \pi^\pm K^\mp$}

\newcommand{\Btohhh}{$B^{\pm}\rightarrow h^{\pm}h^+h^-$}

\newcommand{\Btokpipi}{$B^{\pm}\rightarrow K^{\pm}\pi^+\pi^-$}
\newcommand{\mBtokpipi}{B^{\pm}\rightarrow K^{\pm}\pi^+\pi^-}
\newcommand{\Btokkk}{$B^{\pm}\rightarrow K^{\pm}K^+K^-$}
\newcommand{\mBtokkk}{B^{\pm}\rightarrow K^{\pm}K^+K^-}
\newcommand{\Btopipipi}{$B^{\pm}\rightarrow \pi^{\pm}\pi^+\pi^-$}
\newcommand{\mBtopipipi}{B^{\pm}\rightarrow \pi^{\pm}\pi^+\pi^-}
\newcommand{\Btopikk}{$B^{\pm}\rightarrow \pi^{\pm}K^+K^-$}

\newcommand{\Btojpsik}{$B^{\pm}\rightarrow J\psi K^{\pm}$}

\newcommand{\Btopph}{$B^{\pm}\rightarrow p \bar{p} h^{\pm}$}

\newcommand{\Btoppk}{$B^{\pm}\rightarrow p \bar{p} K^{\pm}$}

\newcommand{\Btopppi}{$B^{\pm}\rightarrow p \bar{p} \pi^{\pm}$}

\newcommand{\Btophikstardetail}{$B^0\rightarrow \phi K^*(892)^0$}

\newcommand{\Btophikstar}{$B^0\rightarrow \phi K^*$}

\newcommand{\Btojpsikstar}{$B^0\rightarrow J/\psi K^*$}
\newcommand{\mBtojpsikstar}{B^0\rightarrow J/\psi K^*}

\mathtoolsset{showonlyrefs}  

\begin{document}
\begin{titlepage}
\pubblock

\vfill
\Title{Review of direct CP violation in two and three body B decays at LHCb}
\vfill
\Author{ Marc Grabalosa G\'andara\support}
\Address{\clermont}
\vfill
\begin{Abstract}
Charmless B hadrons decays offer rich opportunities to test the Standard Model. CP violation in charmless charged two-body and three-body B decays provides ways to measure the CKM angle $\gamma$ and to search for New Physics. Also, vector-vector final states provide additional interesting observables. Hereby, we present the latest LHCb results on hadronic charmless B decays putting emphasis on the direct CP violation measurements. 
\end{Abstract}
\vfill
\begin{Presented}
8$^{th}$ International Workshop on the CKM Unitarity Triangle (CKM 2014)\\
Vienna, Austria, September 8-12, 2014

\end{Presented}
\vfill
\end{titlepage}
\def\thefootnote{\fnsymbol{footnote}}
\setcounter{footnote}{0}
%


\section{Introduction}

Charmless b-hadron decays are a testing ground for the Standard Model as they have contributions from Tree and Penguin diagrams and CP violation may arise from the interference of both. In particular, the measurement of CP violation observables, as well as branching ratio measurements, can lead to an improvement of the CKM matrix elements.
Here, we report the latest analysis performed on charmless B decays by the LHCb detector~\cite{LHCb}.

\section{CP violation in the charmless $B^0_{(s)}\rightarrow K^\pm \pi^\mp$}

The \Btokpipm~and \Bstokpipm~decays can be used for the measurement of direct CP violation by looking at the so-called CP asymmetry which can be defined looking at the decay rates of the self-tagged modes as

\begin{footnotesize}
\begin{equation}
A_{CP}(\mBtokpi) = \frac{\Gamma(\mantiBtokpi)-\Gamma(\mBtokpi)}{\Gamma(\mantiBtokpi)+\Gamma(\mBtokpi)}.
\end{equation}
\end{footnotesize}
This asymmetry has been recently measured by the LHCb with 1~fb$^{-1}$ of data at center-of-mass energy 7~TeV~\cite{Btokpi}. After an efficient selection which takes into account different optimizations for $B^0$ and $B^0_s$ modes, the signal candidates are used to measure the raw asymmetry which are later corrected for detection and production asymmetries. 
Their measured CP asymmetries are

\vspace{-0.4cm}
\begin{footnotesize}
\begin{eqnarray}
A_{CP}(\mBtokpi) &=& -0.080 \pm 0.007 ({\rm stat}) \pm 0.003 ({\rm syst}) \\
A_{CP}(\mBstokpi) &=& 0.27 \pm 0.04 ({\rm stat}) \pm 0.01 ({\rm syst}) 
\end{eqnarray}
\end{footnotesize}
\noindent being the most precise measurement ($10.5\sigma$) of CP violation in \Btokpipm~and the first observation ($6.5\sigma$) of CP violation in $B^0_s$ decays.

\section{CP violation on \Btohhh}

Charmless three-body decays are dominated by processes involving intermediate resonances, and thus, rich interference patterns may arise. After the first evidence of CP violation in \Btohhh~\cite{Btohhh1fba,Btohhh1fbb}, an update of the analysis is performed with 3~fb$^{-1}$~\cite{Btohhh3fb} with the aim of measuring CP violation inclusively but also in the phase space by looking for CP violation asymmetries in local regions of the Dalitz plot. 

New selection criteria including a multivariate technique and new particle identification variables are used. The signal candidates are extracted from an unbinned maximum likelihood fit to the mass spectra of the selected candidates: \Btokpipi, \Btokkk, \Btopipipi~and \Btopikk. 
After including the effects induced by the detector and for the \textit{B} meson production asymmetry and correcting for the acceptance to take into account the non uniformity of the efficiencies, the CP violation asymmetries were measured to be

\begin{figure}[htb!]
\centering
\includegraphics[scale=0.35]{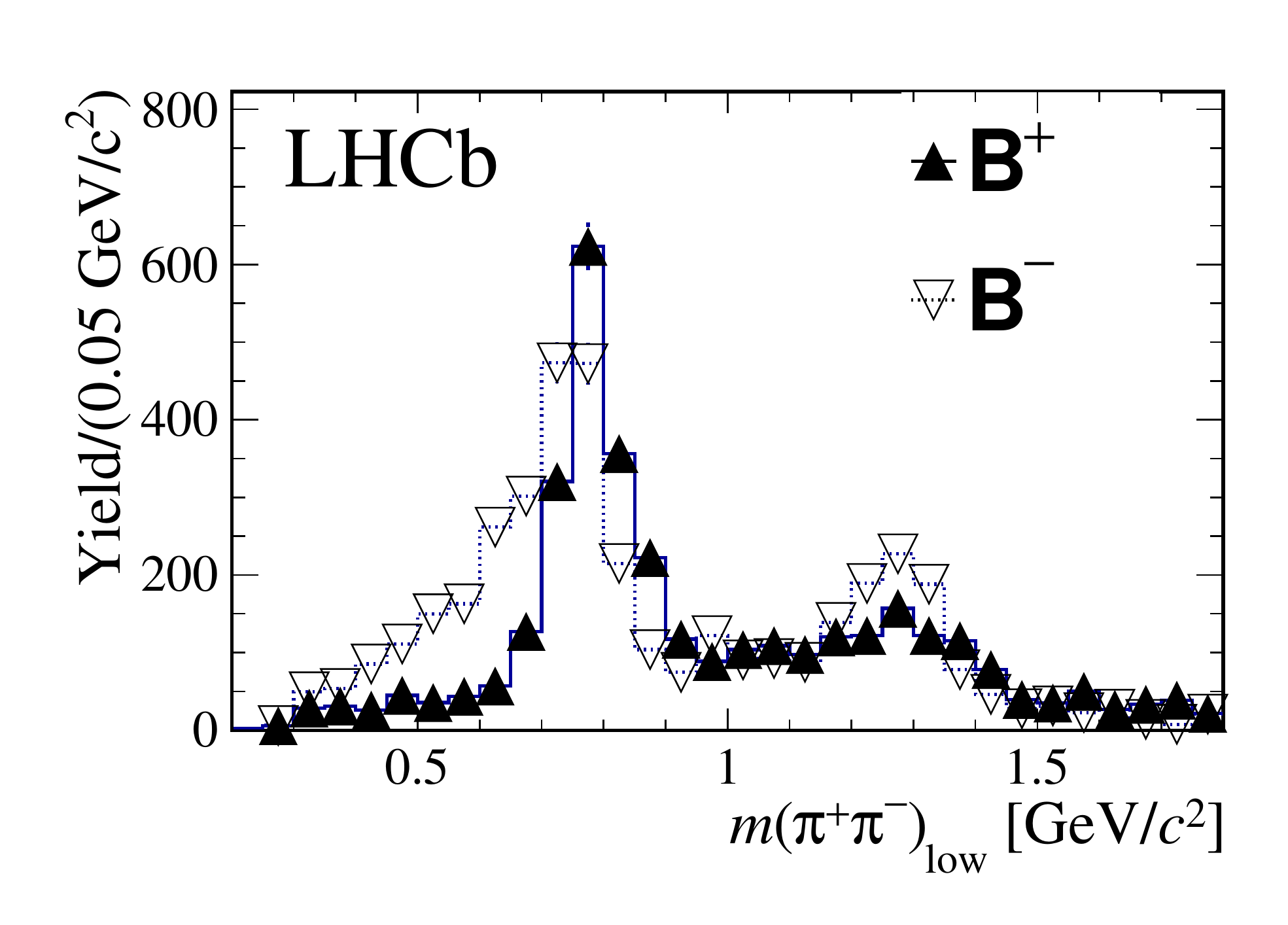}
\includegraphics[scale=0.35]{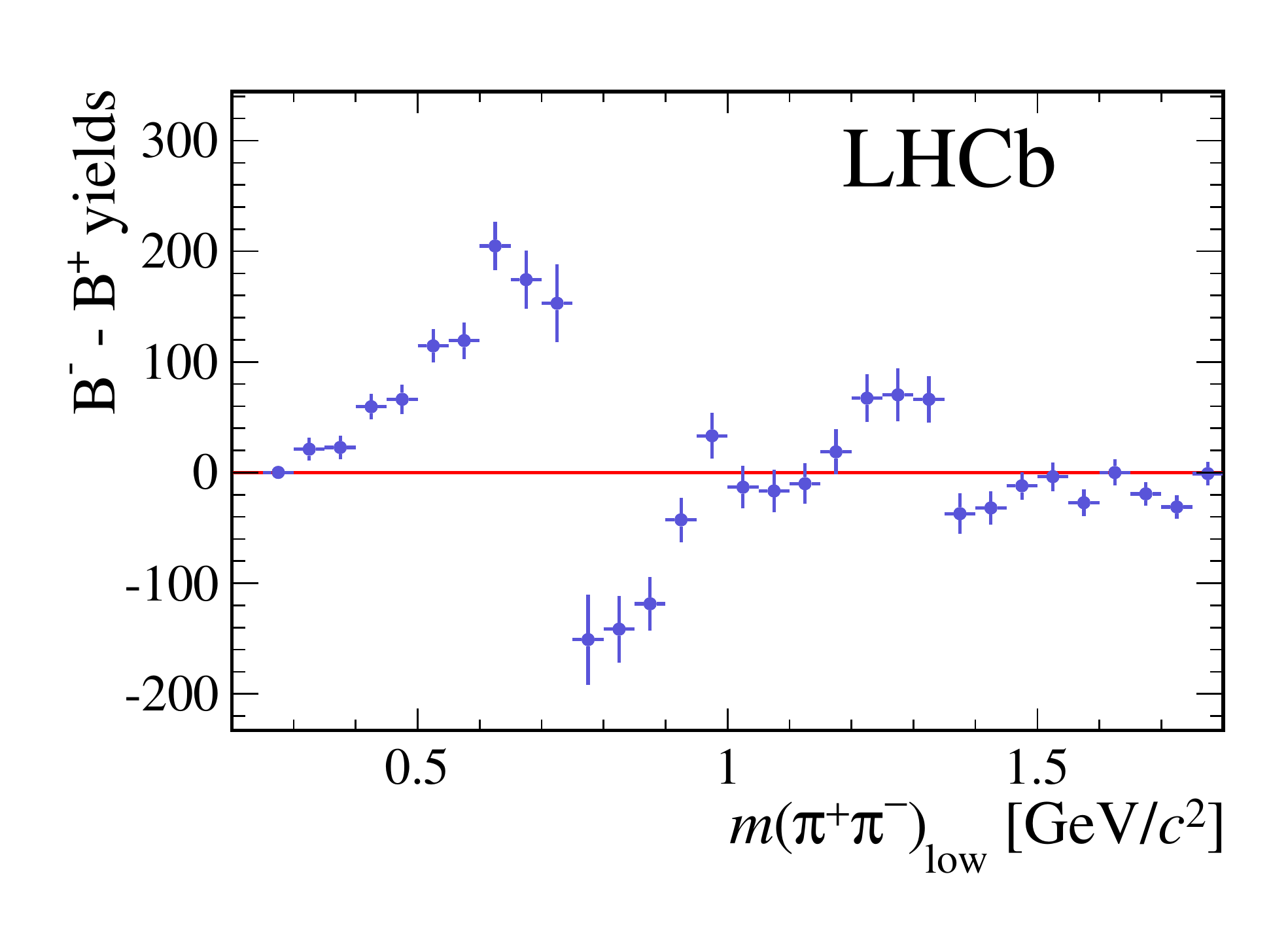}
\caption{{\small Yields and asymmetries for \Btopipipi when $\cos \theta_p < 0$ as a function of m$_{\pi\pi}$.}}
\label{fig:btohhh}
\end{figure}

\vspace{-0.4cm}
\begin{footnotesize}
\begin{eqnarray}
A_{CP}(\mBtokpipi) &=& +0.025 \pm 0.004 ({\rm stat}) \pm 0.004 ({\rm syst}) \pm 0.007 (A_{CP}(J/\psi K)) \\ 
A_{CP}(\mBtokkk) &=& -0.036 \pm 0.004 ({\rm stat}) \pm 0.002 ({\rm syst}) \pm 0.007 (A_{CP}(J/\psi K)) \\
A_{CP}(\mBtopipipi) &=& +0.058 \pm 0.008 ({\rm stat}) \pm 0.009 ({\rm syst}) \pm 0.007 (A_{CP}(J/\psi K)) \\ 
A_{CP}(\mBtokkk) &=& -0.123 \pm 0.017 ({\rm stat}) \pm 0.012 ({\rm syst}) \pm 0.007 (A_{CP}(J/\psi K)). 
\end{eqnarray}
\end{footnotesize}
To study the CP asymmetries in local regions of the Dalitz plots, an adaptative binning to keep the same events per bin was developed; the background was substracted using sWeights~\cite{sweights} and corrections were applied to take into account acceptance effects. Large local raw asymmetries in certain regions of the phase space were observed. The projections on the m$_{hh}$ divided according to the sign of the cosine of the angle between the momenta of the unpair hadron and the resonant daughter with the same-sign charge ($\theta_p$) were also studied, as seen in Figure~\ref{fig:btohhh}.

The charge asymmetry sign flip near the resonance seems to indicate a dominance of the long-distance interference. 
In addition, negative and positive charge asymmetries are also found in the rescattering ($\pi\pi \leftrightarrow KK$) region defined between $1 < (m_{\pi\pi}$ or $m_{KK}) < 1.5$~GeV/c$^2$.

\section{First evidence of CPV in baryonic B decays}

The \Btopph~decays are studied to look for CP violation in baryonic modes on the full 3~fb$^{-1}$ sample produced in LHCb run I~\cite{Btopph}. 
The yields are extracted with an unbinned maximum likelihood fit
to the \Btoppk~and \Btopppi~candidates. 
The Dalitz plots were studied after substracting the background using sWeights and after correcting for acceptance effects, observing clear evidences of the charmonium resonances ($J/\psi, \psi(2S)$ and $\eta_c$), an enhancement at low m$_{p\bar{p}}^2$ and some hints of $\Lambda(1520) \rightarrow pK^-$, for which the branching ratio was computed using \Btojpsik~as control channel, obtaining

\vspace{-0.4cm}
\begin{footnotesize}
\begin{equation}
BR(B^{\pm} \rightarrow \bar{\Lambda}(1520) (\rightarrow \bar{p}K^-) p )= (3.15 \pm 0.48 ({\rm stat}) \pm 0.07 ({\rm syst}) \pm 0.26 ({\rm \mBtojpsikstar})) \times 10^{-7}.
\end{equation}
\end{footnotesize}
The \Btopph~dynamics are studied below the charmonium threshold region (m$_{p\bar{p}} <$ 2.85~GeV/c$^2$), the acceptance-corrected distribution of the cosine of the helicity angle $\theta_p$ shows opposite behaviour of the two modes, which can be generated by non-resonant scattering. This distribution can be used to compute the forward backward asymmetry

\begin{figure}[htb!]
\centering
\includegraphics[scale=0.35]{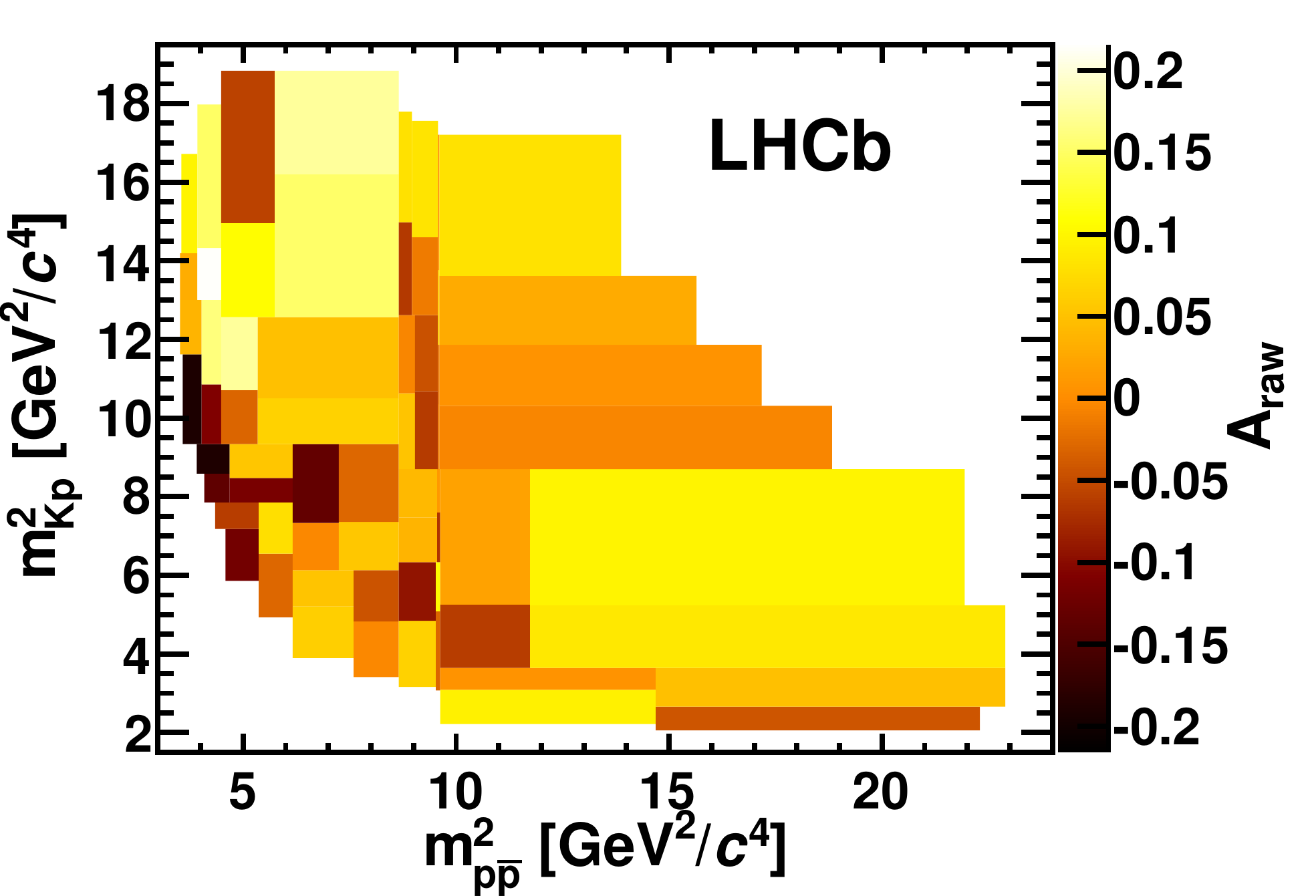}
\includegraphics[scale=0.35]{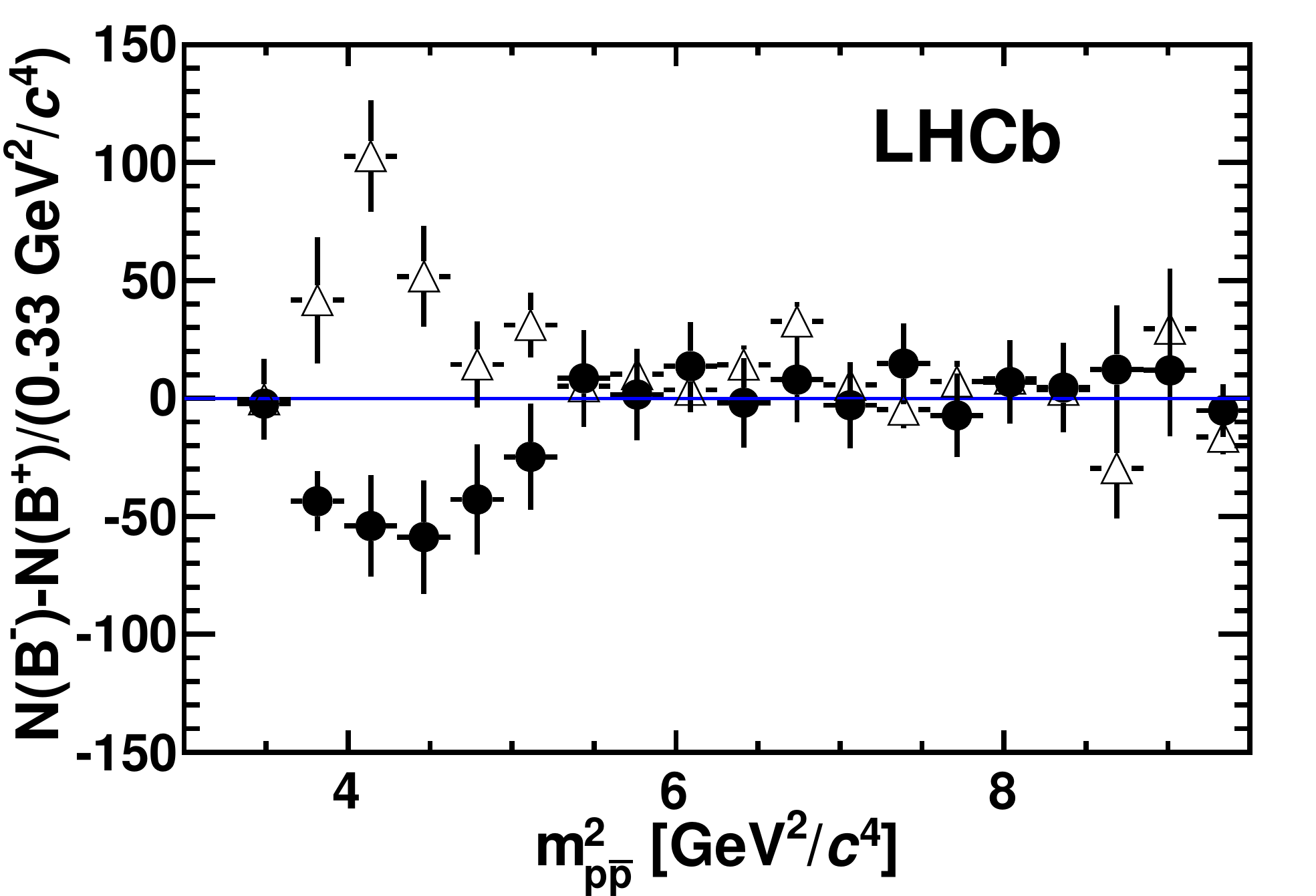}
\caption{\small{Dalitz plot and CP raw asymmetry for \Btoppk.}}
\label{fig:btopph}
\end{figure}

\vspace{-0.4cm}
\begin{footnotesize}
\begin{eqnarray}
A_{FB}(p\bar{p}K^{\pm} < 2.85~{\rm GeV/c^2}) &=& +0.495 \pm 0.012 ({\rm stat}) \pm 0.007 ({\rm syst}) \\
A_{FB}(p\bar{p}\pi^{\pm} < 2.85~{\rm GeV/c^2}) &=& -0.409 \pm 0.033 ({\rm stat}) \pm 0.006 ({\rm syst}). 
\end{eqnarray}
\end{footnotesize}
For \Btoppk~the Dalitz plane was obtained, observing
negative and positive CP asymmetries in different regions of the phase space as seen in Figure~\ref{fig:btopph}. 

For the region m$_{p\bar{p}} <$ 2.85~GeV/c$^2$ and $m_{Kp}^2 >$ 10~GeV$^2$/c$^4$ a 4$\sigma$ evidence of CPV is found, being the first evidence of CP violation in baryonic B decays. The CP asymmetry is measured to be

\vspace{-0.4cm}
\begin{footnotesize}
\begin{eqnarray}
A_{CP}(p\bar{p}K^{\pm}, m_{p\bar{p}} < 2.85~{\rm GeV/c^2}, m_{Kp}^2 > 10~{\rm GeV^2/c^4}) = 0.096 \pm 0.024 ({\rm stat}) \pm 0.004 ({\rm syst}).
\end{eqnarray}
\end{footnotesize}

\vspace{-0.7cm}
\section{Polarization and CP asymmetries on \Btophikstar}

According to the SM, the \Btophikstardetail~proceeds mainly via a gluonic penguin diagram, and thus, it is sensitive to new physics as new contribution may arise inside the loop. 

A simultaneous mass and angular analysis is performed on the \Btophikstardetail \linebreak
and their decay products with 1~fb$^{-1}$ of data~\cite{Btophikstar}. Polarization amplitudes and phases are measured being longitudinal and transverse components of roughly equal amplitudes and their CP asymmetries compatible with zero as shown in Table~\ref{tab:Btophikstar}. 

\begin{table}[htb!]
\begin{footnotesize}
\centering
\begin{subtable}{.5\textwidth}
\centering
\begin{tabular}{l|c}  
Parameter &  LHCb \\
\hline
 $f_L$                     &   0.497 $\pm$ 0.019 $\pm$ 0.015 \\
 $f_{\perp}$               &   0.221 $\pm$ 0.016 $\pm$ 0.013 \\
 $\delta_{\perp}$          &   2.633 $\pm$ 0.062 $\pm$ 0.037 \\
 $\delta_{\parallel}$      &   2.562 $\pm$ 0.069 $\pm$ 0.040 \\
\end{tabular}
\end{subtable}%
\begin{subtable}{.5\textwidth}
\centering
\begin{tabular}{l|c}  
CP parameter &  LHCb \\
\hline
 $A_{CP}^0$                &  -0.003 $\pm$ 0.038 $\pm$ 0.005 \\
 $A_{CP}^{\perp}$          &  +0.047 $\pm$ 0.072 $\pm$ 0.009 \\
 $\delta_{CP}^{\perp}$     &  +0.062 $\pm$ 0.062 $\pm$ 0.006 \\
 $\delta_{CP}^{\parallel}$ &  +0.045 $\pm$ 0.068 $\pm$ 0.015 \\ 
\end{tabular}
\end{subtable}
\caption{{\small Parameters measured in the angular analysis. The first and second uncertainties are statistical and systematic, respectively.}}
\label{tab:Btophikstar}
\end{footnotesize}
\end{table}

The direct CP asymmetry rate is measured with respect to the \Btojpsikstar~channel. $\Delta A_{CP} = (+1.5 \pm 3.2 \pm 0.5) \%$ is consistent with zero and the most precise measurement up to date.

\section{Conclusions}

A short review on the latest LHCb results on charmless 2- and 3-body decays is presented. The 
most precise CP violation in \Btokpipm~has been obtained as well as the first evidence of CP violation in $B^0_s$ decays. In the \Btohhh~and \Btoppk~channels, a large CP asymmetry has been found in regions of the Dalitz which do not correspond to resonant contributions, including the first evidence of CP violation in baryonic B decays. 
For the \Btophikstardetail~equal longitudinal and transversal polarization are obtained with CP asymmetries compatible with zero. 



\end{document}